\documentclass[a4paper,11pt]{revtex4}
\usepackage{graphicx,amsmath,amssymb,bm,latexsym,color,epsf}
\makeatletter
\@addtoreset{equation}{section}

\makeatother
\newcommand{\be}{\begin{equation}}
\newcommand{\ee}{\end{equation}}
\newcommand{\bea}{\begin{eqnarray}}
\newcommand{\eea}{\end{eqnarray}}

\newcommand{\nn}{\nonumber}

\newcommand{\bR}{\bar R}

\newcommand{\bnabla}{\bar\nabla}

\newcommand{\bg}{\bar g}

\newcommand{\tth}{{h^{TT}}}

\newcommand{\bphi}{\bar\phi}

\newcommand{\rad}{\psi}

\begin{document}

\begin{titlepage}

\title{Asymptotic safety in $O(N)$ scalar models coupled to gravity}

\author{Peter Labus}
\email{plabus@sissa.it}
\affiliation{
SISSA, via Bonomea 265, I-34136 Trieste
\\
and 
INFN, Sezione di Trieste, Italy
}
\author{Roberto Percacci}
\email{percacci@sissa.it}
\affiliation{
SISSA, via Bonomea 265, I-34136 Trieste
\\
and 
INFN, Sezione di Trieste, Italy
}
\author{Gian Paolo Vacca}
\email{vacca@bo.infn.it}
\affiliation{INFN, Sezione di Bologna,
via Irnerio 46, I-40126 Bologna}

\begin{abstract}
We extend recent results on scalar-tensor theories
to the case of an $O(N)$-invariant multiplet.
Some exact fixed point solutions of the RG flow equations are discussed.
We find that also in the functional context, on employing a 
standard ``type-I'' cutoff, 
too many scalars destroy the gravitational fixed point.
For $d=3$ we show the existence of the gravitationally dressed 
Wilson-Fisher fixed point also for $N>1$.
We discuss also the results of the analysis for a different, scalar-free, coarse-graining scheme.
 
\end{abstract} 

\maketitle

\end{titlepage}
\newpage
\setcounter{page}{2}

\section{Introduction}
In the last years many efforts have been devoted to studying the possibility of defining UV complete QFTs
which may describe gravitational interactions, 
possibly in the presence of other matter fields.
Such a program is defined within the paradigm of Asymptotic Safety~\cite{W1} and can be seen as a bottom-up approach which seeks a consistent extension
to UV of results of low energy effective field theories.
It is related to the construction of the non perturbative RG flow of the effective action, which should show a UV interacting fixed point (in theory space) with a finite number of relevant directions.

In a recent paper \cite{pv1} two of us discussed 
fixed-functional solutions
in a scalar-tensor theory with action containing a potential $V(\phi)$
and a generic non-minimal interaction $-F(\phi)R$.
These functions were subjected to a renormalization group flow
depending on a cutoff $k$.
It was found that in generic dimension $d$ the flow equations for the dimensionless
functions $v(\varphi)=k^{-d}V(\phi)$ and $f(\varphi)=k^{2-d}F(\phi)$
of the dimensionless field $\varphi=k^{1-d/2}\phi$
admit some simple exact solutions where $v=v_0$ is constant and
$f=f_0+f_2\varphi^2$.
These solutions can have interesting applications to cosmology in $d=4$.
\cite{hprw}.
Unfortunately it seems that one of these solutions,
with $f_0$ and $f_2$ both nonzero, does not survive,
at least in the same form, when one applies the
``renormalization group improvement'',
while another, which remains unchanged by the
``renormalization group improvement'',
has $f_0=0$ but $f_2<0$, making its physical relevance questionable.

In this paper we consider the case when 
there are $N$ scalar fields, with an $O(N)$-invariant
action of the general form
\be
\label{action}
\int d^dx\sqrt{g}\left(V(\phi)
+\frac{1}{2}\sum_a(\nabla\phi^a)^2-F(\phi)R\right)\ .
\ee
where now $V$ and $F$ are functions of the radial degree of freedom
and we use the notation $\phi=\sqrt{\sum_a \phi^a\phi^a}$ and for the dimensionless rescaled field
$\varphi=\sqrt{\sum_a \varphi^a\varphi^a}$.
(In section IV it will prove more convenient to think of
$v$ and $f$ as functions of the variable $\rho=\varphi^2/2$.)
The remaining $N-1$ angular degrees
of freedom will be referred to as the Goldstone bosons.
One of the main results of this note is to show that in the case
when $N>1$ there exists also another solution with $f_0=0$ and $f_2>0$.

The other main result will be the confirmation, 
on employing a standard cutoff of ``type-I''
(in the terminology of \cite{cpr2}),
of the existence  of an upper bound on the number of scalars in order to obtain a FP with positive $f$.
Such bounds have been discussed earlier but only for minimally 
coupled fields.
We will see that the existence of the potential and of the nonminimal interactions does not substantially change the picture.

In the same scheme for $d=3$ we also discuss the existence 
of a non trivial gravitationally dressed fixed point 
and give a specific solution for $N=2$.
Finally we will employ an alternative coarse-graining scheme
in order to test the scheme-dependence of the analytical scaling solutions.

\section{Flow equations}

For the construction of the flow equations we follow the same steps
as in \cite{pv1}, which we briefly recall.
The calculation is based on the background field method,
but instead of the usual linear classical-background-plus-quantum-fluctuation split we use an exponential parametrization
for the metric of the form 
$g_{\mu\nu}=\bar g_{\mu\rho}(e^h)^\rho{}_\nu$.
This has some concrete practical advantages that will be 
mentioned later, 
but the original theoretical motivation is that it
respects the non-linear nature of the metric.
The fluctuation field is further decomposed into its irreducible
spin-two, spin-one and spin-zero components $h_{\mu\nu}^{TT}$,
$\xi_\mu$, $\sigma$ and $h$.
Then one calculates the second variation of the action with
respect to the fields, which appears as a central ingredient
in the flow equation.
In order to have a diagonal Hessian we define new scalar fields
that involve a mixture of $\sigma$ and $\phi$.
We then choose a very simple ``unimodular physical gauge''
which amounts simply to suppressing the fields $\xi$ and $h$.
We describe briefly these steps in the Appendix. 
Finally we introduce a ``type I'' cutoff, which amounts to replacing
all occurrences of $-\bnabla^2$ by $-\bnabla^2+R_k(-\bnabla^2)$
\cite{cpr2}. 
In particular we will use the cutoff profile function 
$R_k(z)=\left(k^2-z\right)\theta\left(k^2-z\right)$~\cite{Litim}.
Note that this cutoff is ``spectrally adjusted'',
in the sense that it depends on some running couplings,
and also depends explicitly on the background scalar field,
in addition to the background metric.
Later we shall consider also other types of cutoff.

In general dimension $d$ and for any number of scalars,
denoting by a dot the derivative with respect to RG time $t=\log k$,
the full flow equations are
\bea
\label{flowvfull}
\dot v&=&-d v+\frac{1}{2} (d-2) \varphi\,  v' +c_d \frac{(d-1) \left(d^2\!-\!d\!-\!3\right)}{d+2}+
c_d \frac{(d-2) (d+1) \left(2 \dot f-(d-2) \varphi  f'\right)}{4 \left(d+2\right) f}\\
&{}&\!\!\!\!+c_d\frac{2 \left(d^2\!-\!4\right) \!f\!+\!(d\!-\!2) \left(1\!+\!v''\right) \left(2 \dot f\!-\!6 f\!-\!(d\!-\!2) \varphi  f'\right)\!+\!
4 (d\!-\!1) f' \left(2 \dot f'\!+\!(d\!-\!1) f'\!-\!(d\!-\!2) \varphi  f''\right)}{2 (d+2) \left(2 (d-1) \left(f'\right)^2+(d-2) f \left(1+v''\right)\right)}
\nn
\\
&&
+c_d\frac{(N-1)\varphi}{\varphi+v'(\varphi)}
\nonumber
\eea
%
%
\bea
 \label{flowffull}
\dot f &=&(2-d) f+\frac{1}{2} (d-2) \phi  f'-c_d\frac{d^6-2 d^5-15 d^4-46 d^3+38 d^2+96 d-24}{12 \left(d+2\right) (d-1) d}\\
&{}&\!\!\!\!-c_d\frac{\left(d^5\!-\!17 d^3\!-\!60 d^2\!+\!4 d\!+\!48\right) \left(2 \dot f\!-\!(d\!-\!2) \varphi  f'\right)}{48 (d-1) d (d+2) f} -c_d \left( (d-2) f f''+2 f'{}^2\right) \times \nonumber\\
&{}&\!\!\!\!\times \frac{(d-2) (d+2) f^2+(d-1)f'{}^2 \left((d-2) \phi  f' \!-\!2 \dot f\right)+2 (d\!-\!1) f f' \left((2-d) \varphi  f''\!+\!(d+2) f'\!+\!2
   \dot f'\right)}{(d+2) f \left(2 (d-1) \left(f'\right)^2+(d-2) f \left(v''+1\right)\right)^2} 
   \nonumber\\
   &{}&\!\!\!\!+c_d \frac{f' \left((d-2) \varphi  \left(4 (d-1) f''+(d-2) \left(v''+1\right)\right)-8 (d-1) \dot f'\right)+2 (d-2)  \left( d\, f v''-\dot f
   \left(v''+1\right)\right)}{24 \left(2 (d-1) \left(f'\right)^2+(d-2) f \left(v''+1\right)\right)}\nonumber
   \\
&&
-c_d\frac{d(N-1)\varphi}{12(\varphi+v'(\varphi))}
-c_d\frac{(N-1)\varphi f'(\varphi)}{(\varphi+v'(\varphi))^2}
\eea
where $c_d^{-1}=(4\pi)^{d/2} \Gamma(d/2+1)$.
The only difference between the case $N=1$ discussed in \cite{pv1} 
and the case of general $N$ is the addition to the beta functionals
of $v$ and $f$ of the contribution of the Goldstone modes,
which can be easily identified by the factor $N-1$.
The ``RG-unimproved'' or ``one-loop''
flow equations can be obtained by replacing in the r.h.s.
\be
\dot f\to -(d-2)f+\frac{d-2}{2}\varphi f'\ ;
\qquad
\dot f'\to -\frac{d-2}{2}f'+\frac{d-2}{2}\varphi f''\ .
\ee
which is equivalent to setting $\dot F$ and $\dot F'$
to zero in the equation for the dimensionful functions $V$ and $F$.

\section{Scaling solutions}

We list here some simple analytic solutions 
that exist in any dimension.
Making the ansatz that $v$ and $f$ are both constant leads
to a fixed point FP1.
Using the simple, unimproved equations, it has the following coordinates:
\bea
v_*&=&c_d\left[\frac{(d-1)(d-2)}{2d}+\frac{N-1}{d}\right]\ ,
\\
f_*&=&
-c_d\frac{d^5-4d^4-7d^3-50d^2+60d+24}{24d(d-1)(d-2)}
-c_d\frac{(N-1)d}{12(d-2)}\ .
\eea
We note that the first fraction in $f_*$ is positive for $d<6.17$.
Thus for $N=1$ this is an upper bound on the dimension
dictated by positivity of Newton's constant.
Having additional scalars lowers this bound.
The bound becomes lower than four between $N=11$ and $N=12$.
Thus, the fixed point has negative Newton's constant
when there are more than $11$ scalars, a result that is in
rough agreement with previous calculations \cite{dep}
that also used a similar cutoff.

If we include the RG improvement the coordinates of FP1 change to:
\bea
v_*&=&c_d\left[\frac{(d^2-1)(d-2)}{d(d+2)}+\frac{N-1}{d}\right]\ ,
\\
f_*&=&
-c_d\frac{d^6-2d^5-15d^4-46d^3+38d^2+96d-24}{12d(d-1)(d^2-4)}
-c_d\frac{(N-1)d}{12(d-2)}\ .
\eea
Now, for $N=1$ positivity of Newton's constant requires $d<5.73$,
and the bound becomes lower than four between $N=14$ and $N=15$.
Thus, the fixed point has negative Newton's constant
when there are more than $14$ scalars.

If we make the ansatz that $v$ is constant and $f$ is of the
form $f_0+f_2\varphi^2$, there is a solution FP2
for the unimproved fixed point equations
\bea
v_*&=&c_d\left[\frac{(d-1) (d-2)}{2 d}+\frac{N-1}{d}\right]\ ,
\\ 
f_*(\varphi)&=&
c_d\left[\frac{d^5-4 d^4-7 d^3-50 d^2+84 d+24}{24 d (d-1) (d-2)}
-\frac{(N-1)(d^2-d+12)}{12(d-1)(d-2)}\right]
+\frac{\varphi^2}{2(d-1)}\,.
\eea
We observe that while $f_2$ is always positive,
$f_0$ becomes negative when either $d$ or $N$ become
too large. For example for fixed $N=2$ this happens at $d\approx 5.8$
and for fixed $d=4$ this happens for $N\approx 5.6$.
The solution is probably unphysical in these cases.

The same ansatz does not yield a solution for the improved equations.
We suspect that there may be a solution
with very similar properties but different functional form.
The search of such generalization is beyond the scope of this paper.

If we make the ansatz that $v$ is constant and $f$ is proportional
to $\varphi^2$ (i.e. that $f_0=0$), the fixed point equation is quadratic and admits two
real solutions FP3 and FP4.
They are the same for the improved and unimproved flow equations.
The expressions for arbitrary $d$ are quite long, so
we only give here the formulae for $d=3$
\be
v_*=\frac{N}{18\pi^2}\ ,
\qquad
f_*=-\frac{9N-80\pm\sqrt{9N^2-264N+5296}}{96(N-1)}\varphi^2
\ee
and $d=4$
\be
v_*=\frac{2+N}{128\pi^2}\ ,
\qquad
f_*=-\frac{6N-41\pm\sqrt{4N^2-100N+1321}}{48(N-1)}\varphi^2\ .
\ee
where the upper sign corresponds to FP3 
and the lower sign corresponds to FP4.
We note that FP3 has a finite $N\to1$ limit,
while in the case of FP4 there is a divergence.
In fact for $N=1$ the fixed point FP3 had already been seen
in \cite{pv1}, but FP4 does not exist in that case.
Imposing that $f>0$ we find that FP3 is always unphysical
while FP4 is acceptable for $0<N<15.33$ in $d=3$ 
and $1<N<11.25$ in $d=4$.

\section{Large-$N$ limit}

All the solutions described in the previous section
have the property that the function $f_*$ is not
positive if the number of scalars exceeds a certain limit.
In order to confirm that this behavior persists
we consider here the large-$N$ limit of the flow.
It is very convenient to change variable and use $\rho=\varphi^2/2$
as the argument of the functions $v$ and $f$.
It is easy to see that in the large-$N$ limit both the potentials $v$ and $f$ as well as $\rho$ scale linearly in $N$.
In particular in such a limit the flow equations, in terms of all the quantities already rescaled by $N$, read simply
\bea
\dot v &=&-d v + (d-2) \rho  \,v'+\frac{c_d}{1+v'}\nonumber \\
\dot f &=& -(d-2) f+\left((d-2) \rho -\frac{c_d}{\left(1+v'\right)^2}\right) f'  -\frac{ d}{12}\frac{c_d} {1+v'}
\label{eqlargeN}
\eea
The fixed point equation for the scalar potential $v$ is the same as in flat space. There is a solution with a constant potential $v$ and one with a nontrivial one which is known analytically \cite{marchais}. 
In the first case the solution to Eqs.~(\ref{eqlargeN}) is given by
\be
v=\frac{c_d}{d} \quad,\quad  f=-\left(\frac{d}{12(d-2)} c_d+2 a\right) +  2a \frac{d-2}{c_d} \rho \,,
\ee
where $a$ is a constant of integration. For such a line of fixed points it is clear that, for $d>2$  and for any value of $a$, the function $f$ is unphysical since either the constant part or the coefficient of $\rho$ are negative.

In order to deal with the  second case with the non trivial solution for $v$, its equation is best written, 
for $d$ not an even integer, in terms of $w(\rho)=v'(\rho)$ in the implicit form
\be
\rho=c_d \frac{d}{4}  \ _2F_1\left(2,1-\frac{d}{2};2-\frac{d}{2};-w\right)
\ee
Therefore the scalar sector presents the Wilson Fisher type global scaling solution for $2<d<4$.
Also the fixed point equation for $f$ can be further simplified in terms of $w$ and defining a new function $g(w)=f(\rho)$ one has
\be
0=-(d-2) g+2 w g'-\frac{ d}{12}\frac{c_d} {1+w}
\ee
which can be solved analytically and gives the implicit solution:
\be
f(\rho)=g(w)=-\frac{d \,c_d}{12 (d-2)} \, _2F_1\left(1,1-\frac{d}{2};2-\frac{d}{2};-w\right)
\ee
which is always negative for $d>2$ and therefore such a fixed point for $2<d<4$ appears in the large $N$ limit to lead to negative gravitational interactions in the current formulation.
We also note that in the asymptotic region $\rho\to\infty$ one has $f(\rho)= -\frac{2 }{3 d (d-2) } \rho$.

\section{Stability analysis}

We discuss here the linearization of the flow in $d=4$ around the
fixed point FP1.
We begin with the simpler ``RG-unimproved'' case,
which yields the following linearized equations:
\bea
0&=& 
-(\lambda +4)\delta v 
+\left(\varphi-\frac{N-1}{32\pi^2\varphi}\right)\delta v'
-\frac{1}{32\pi^2}\delta v''  \\
0&=& -(\lambda +2)\delta f
+\left(\varphi-\frac{N-1}{32\pi^2\varphi}\right)\delta f'
+\frac{N-1}{96\pi^2\varphi}\delta v'
+\frac{1}{96\pi^2}\delta v''
-\frac{1}{32\pi^2}\delta f'
\eea
In this case the critical exponents are equal to their classical values:
\bea
&{}&\theta_1=4 , \qquad \qquad  \qquad \ w_1^t=(\delta u,\delta\! f)_1 = (1,0) \\
&{}&\theta_3=2 , \qquad  \qquad  \qquad \ w_3^t=(\delta u,\delta\! f)_3 = (0,1) \nonumber\\
&{}&\theta_5=0 , \qquad  \qquad \qquad \ w_5^t=(\delta u,\delta\! f)_5 
=\left(0, \nonumber-\frac{N}{32\pi^2}+\varphi^2\right)
\eea

The solution FP1 for the full fixed point equations in $d=4$
has coordinates 
$\left(v^*=\frac{N+4}{128 \pi ^2},
f^*=\frac{169-12 N}{2304 \pi ^2}\right)$.
Linearizing the full flow equations around this solution
we get the eigenvalue equations for the eigenperturbations:
\bea
0&=& 
-(\lambda +4)\delta v 
+\frac{72\lambda}{169\!-\!12 N}\delta \!f
+ \left(\varphi-\frac{N-1}{32\pi^2\varphi}\right)\delta v'
+\frac{72\varphi}{12N\!-\!169}\delta\!f'-\frac{1}{32\pi^2}\delta v''  \\
0&=& \left(\frac{3\lambda(45-4N)}{12N\!-\!169}-2\right)\delta\!f
+\left(\frac{3(4N\!-\!45)\varphi}{12N-169}
-\frac{N-1}{32\pi^2\varphi}\right)\delta\!f'
+\frac{N-1}{96\pi^2\varphi}\delta v'
+\frac{1}{96\pi^2}\delta v''
-\frac{1}{32\pi^2}\delta\!f''
\nonumber
\eea

We can study the spectrum of the leading eigenvalues analytically. We find four relevant and one marginal direction for $N<\frac{45}{4}$, while for $\frac{45}{4}<N<\frac{169}{12}$ there are only two relevant and one marginal directions,
since $\theta_2$ and $\theta_4$ become negative. In particular
\bea
&{}&\theta_1=4 , \qquad \qquad  \qquad \ (\delta v,\delta\! f)_1 = (1,0) \\
&{}&\theta_2=2\!+\!\frac{68}{3 (45\!-\!4N)} , \ \  
(\delta v,\delta\! f)_2 = \left(\frac{72}{12 N\!-\!101}, 1\right) \nonumber\\
&{}&\theta_3=2 , \qquad\qquad\qquad\ 
(\delta v,\delta\! f)_3 = 
\left(-\frac{29 N}{544\pi^2}+\varphi^2,
\frac{N(169\!-\!12 N)}{3264\pi^2}\right)
\nonumber\\
&{}&\theta_4=\frac{68}{3 (45\!-\!4N)} ,\quad\quad
(\delta v,\delta\! f)_4
=\bigg(-\frac{3N(48N(3 N\!-\!76)+22475)}{16\pi^2(4N\!-\!45)(6N\!-\!59) (12N\!-\!101)}+\frac{72}{12N\!-\! 101}\varphi^2,
\nn
\\
&&
\qquad\qquad\qquad\qquad\qquad\qquad\qquad 
\qquad\qquad\qquad 
-\frac{N(12N\!-\!169)(12N-125)}
{96\pi^2(4N\!-\!45)(12 N\!-\!101)}+\varphi^2\bigg)
\nonumber\\
&{}&\theta_5=0 , \qquad  \qquad \qquad \ (\delta v,\delta\! f)_5 
=\bigg(\frac{29 N (N+2)}{17408\pi^4}-\frac{29(N+2)}{272\pi ^2}\varphi^2+\varphi^4,
\nn\\
&&
\qquad\qquad\qquad\qquad\qquad\qquad\qquad 
\qquad\qquad\qquad 
\frac{N (N+2) (12 N\!-\!227)}{52224\pi ^4}-\frac{(N+2) (12 N\!-\!169) }{1632\pi^2}\varphi^2\bigg)
\nonumber
\eea
We note that the eigenvalues come in groups
within which they are shifted by two.
This behavior had already been observed and
explained in \cite{narain1}.
%
%
%
%

\section{The gravitationally dressed Wilson-Fisher fixed point}

\begin{figure}
\includegraphics[width=7.5cm]{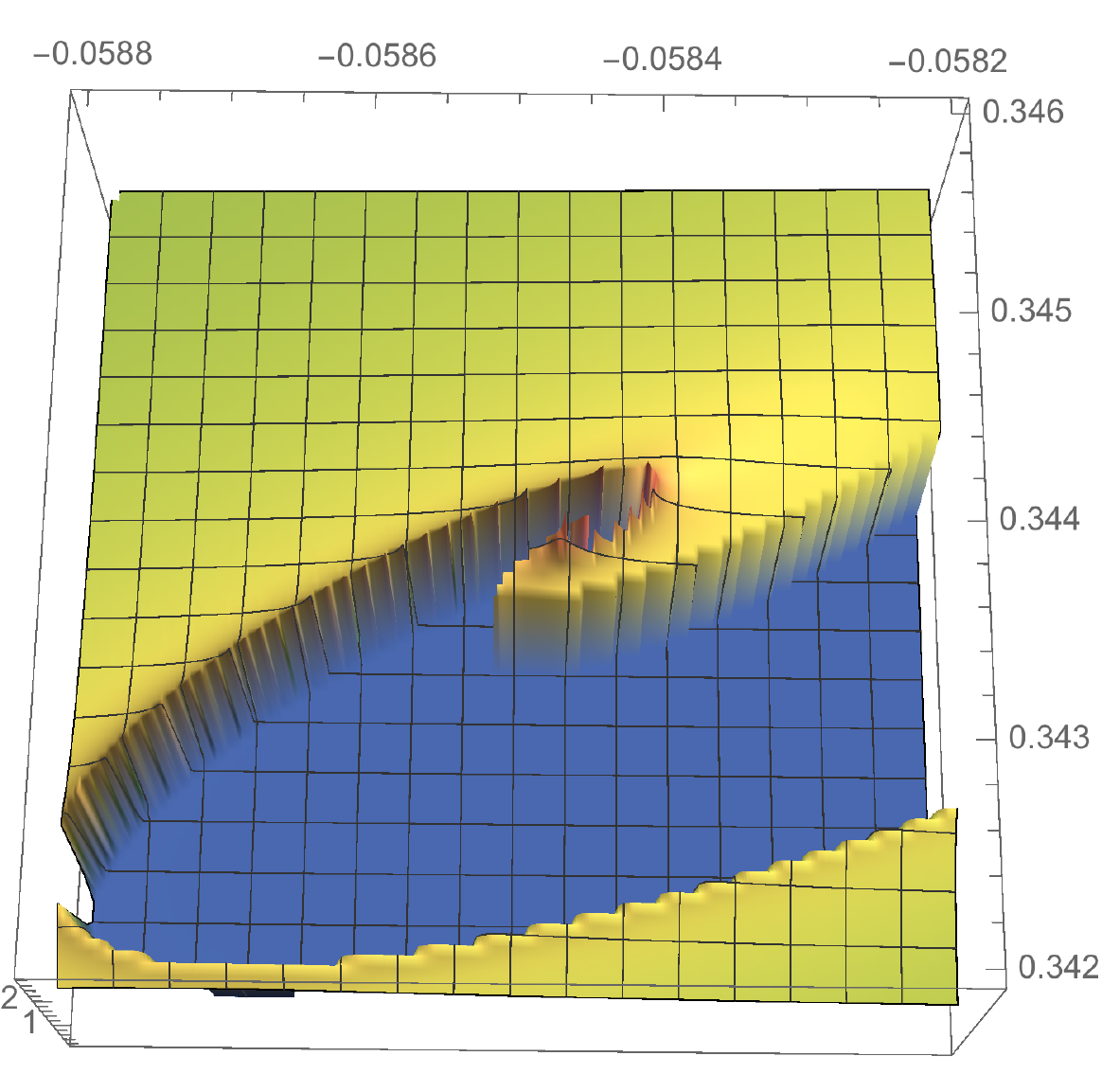}
\
\includegraphics[width=7.5cm]{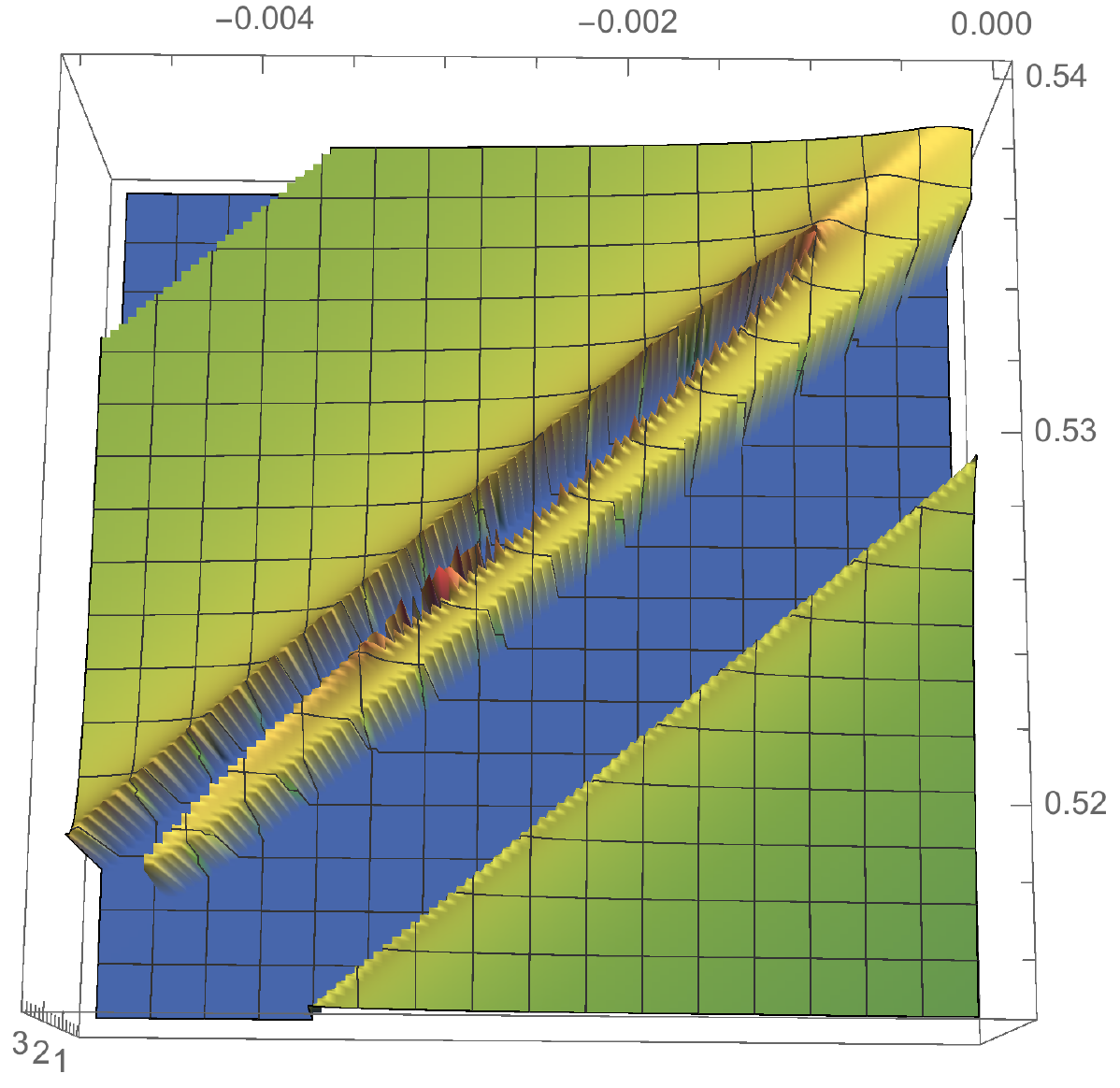}
\caption{The maximal value of the field reached in the numerical
evolution before encountering a singularity, as a function 
of the initial conditions $v''(\varphi=0)$ and $f''(\varphi=0)$,
for the case $d=3$ and $N=1$ (left panel) and $N=2$ (right panel).
A clear spike can be seen in the center of each figure.
}
\label{WFN1N2}
\end{figure}

In \cite{pv1} we looked for scaling solutions in $d=3$
with a potential resembling that of the Wilson-Fisher fixed point. 
We concentrated on the unimproved equations,
and found a solution for sufficiently small $\varphi$ 
whose potential is almost indistinguishable from the
Wilson-Fisher potential, and with a function $f$
that starts out positive at $\phi=0$ but has negative
second derivative, such that it crosses zero at
some critical value $\varphi\approx 0.92$.
The analysis of the solution beyond this critical value
proved hard and we were unable to establish its global existence.
Perhaps more important, the Hessian becomes ill-defined at the
critical point, so that the equations themselves become unreliable.
A little later, using more powerful numerical techniques,
a global solution was found for the improved RG equation \cite{bk}.
The fixed point potential for this solution is again
very similar to the Wilson-Fisher potential,
but the function $f$ now has positive second derivative
and is positive everywhere, avoiding the issues mentioned above.
With hindsight this solution can be seen also with the
simpler techniques used in \cite{pv1}, such as
Taylor expansion and the shooting method.
Near the origin it has the expansion
\bea
v&=&0.00935540-0.0292660 \varphi ^2+0.00359119 \varphi ^4+1.14530 \varphi ^6+\cdots
\nonumber \\
f&=&0.0686040+0.172245 \varphi ^2-0.132631 \varphi ^4+0.390317 \varphi ^6+\cdots
\nonumber
\eea
We have looked for generalizations of this solution for $N>1$. 
Treating $N$ as a continuous variable, candidate scaling solutions
can be found with the shooting method.
We show, as an example, in Fig.~\ref{WFN1N2} the cases $N=1$ and $N=2$. 
A spiraling structure appears close to the fixed point which is characterized by a very sharp relative peak. 

For $N=2$ the solution can be approximated near the origin by
\bea
v&=&0.0146691\, -0.00148208 \varphi ^2-0.438516 \varphi ^4+4.35302 \varphi ^6+\cdots
\nonumber \\
f&=&0.053294\, +0.26337 \varphi ^2-0.570919 \varphi ^4+1.43572 \varphi ^6+\cdots
\nonumber
\eea
The shape of the potential $v$ shows that the scaling solution is characterised by a broken phase.

The asymptotic behavior of the solution 
for large $\varphi$ is
\bea
u_{as}(\varphi)&=&
A \varphi ^6+\frac{23}{1440 \pi ^4 B \varphi ^2}+\frac{240 \pi ^4 B^2 (16 B+5 N-4)-1357 A}{216000 \pi ^6 A B^2 \varphi ^4}+\cdots\nonumber\\
f_{as}(\varphi)&=&
B \varphi ^2+\frac{23}{36 \pi ^2}+\frac{1219}{25920 \pi ^4 B \varphi ^2}
-\frac{71921 A+2160 \pi ^4 B^2 (16 B+5 N-4)}{4665600 \pi ^6 A B^2 \varphi ^4}+\cdots \quad
\eea
For the case $N=2$ 
there is a very good match between the numerical solution found with the shooting method around the origin
and the asymptotic behavior given above for $A\simeq 5.149$ and $B\simeq 0.273$ around $\varphi\simeq0.65$. 
Therefore we consider this as a good candidate for 
a global scaling solution for the gravitationally 
dressed ${\rm O(2)}$ scalar model.

The critical exponents associated to this fixed point ($N=2$) have not been determined with sufficient precision using polynomial methods. We can only report that they are not too far from the values obtained for the $N=1$ case.
More sophisticated numerical methods must be used.

\section{Scalar-free cutoff}

So far we have considered a type-I cutoff which in the gravitational
sector has the general form $F(\bar\phi)R_k(-\bnabla^2)$.
The presence of the prefactor $F$ is useful because the effect
of adding the cutoff results simply in the replacement of
$-\bnabla^2$ by $P_k(\bnabla^2)=-\bnabla^2+R_k(-\bnabla^2)$
in the Hessians.
However, this advantage comes at a price:
the cutoff depends on running couplings
and there is an explicit breaking of the scalar split symmetry
$\bar\phi^a\to\bar\phi^a+\epsilon^a$, 
$\delta\phi^a\to\delta\phi^a-\epsilon^a$.
As discussed in \cite{dietz3}, already in pure scalar theory
the presence of the scalar background field 
in the cutoff action can lead to unphysical results.
This argument casts doubts on the use of these cutoffs,
and it is important to explore also cutoffs that
do not have a prefactor $F$.
In the Einstein-Hilbert truncation, where $F=1/(16\pi G)$,
such cutoffs were called ``pure'' \cite{narain3}. 
Here we shall call them ``scalar-free''. 
It is important to keep in mind that any cutoff still
depends on the background metric, and that this dependence will
have to be dealt with by other means, see e.g. \cite{dietz4,becker}.
Here we only get rid of the $\bar\phi$--dependence and have the scalar split Ward identity automatically fulfilled.
In particular we shall add in the diagonal entries for $h^{TT}_{\,\mu\nu}$ and $\sigma''$ of the Hessian given in Eq.~(\ref{hessfinal})  a cutoff $\gamma\, k^{d-2}(k^2-z)\theta(k^2-z)$, 
with the same sign of the Laplacian term, to implement the coarse-graining procedure. For all the other terms we shall use the more common
$(k^2-z)\theta(k^2-z)$.

We do not report here the form of the flow equations in any dimension,
which contains hypergeometric functions.
For $d=4$ the equations read:
\bea
\dot v&=&-4v+\varphi v'-\frac{1}{8\pi^2}
-\frac{f\left(f(v''+1)\log\left(\frac{3 f'^2}{f(v''+1)}+1\right)-3 f'^2\right)}{144\pi^2f'^4}
\nn\\
&&
+\frac{3\gamma\left(-\gamma^2+4\gamma f-2\gamma(\gamma-2f)\log \left(\frac{f}{\gamma}\right)
-3 f^2\right)}{16 \pi ^2 (\gamma
   -f)^3}+\frac{(N-1) \varphi }{32 \pi ^2 \left(v'+\varphi \right)}
\\
\dot f&=&-2f+\varphi f'+\frac{19}{384\pi^2}
+\frac{f\left(\frac{6f'^2\left(f f''+f'^2\right)}{3f'^2+f\left(v''+1\right)}
-\left(2 f f''+3 f'^2\right) 
\log\left(\frac{3 f'^2}{f(v''+1)}+1\right)\right)}{288 \pi ^2 f'^4}
   \nn\\&&
   +\frac{10 \gamma  (\gamma -2 f) (\gamma -f)-5
   \gamma  \left(\gamma ^2-3 \gamma  f+4 f^2\right) \log \left(\frac{f}{\gamma }\right)}{96
   \pi ^2 (\gamma -f)^3}
\nn\\&&
+\frac{\gamma\left(-\gamma +(\gamma-2f)\log\left(\frac{f}{\gamma}\right)+f\right)}{96\pi^2(\gamma-f)^2}
-\frac{(N-1) \varphi  \left(3 f'+v'(\varphi )+\varphi \right)}{96 \pi ^2 \left(v'+\varphi\right)^2}
\eea
Unlike (\ref{flowvfull},\ref{flowffull}),
these are transcendental equations and cannot be solved analytically.
For a comparison we discuss
only the  scaling solution FP1
which is defined by $v(\varphi)=v_0$ and $f(\varphi)=f_0$.
With this ansatz the fixed point conditions reduce to 
the following algebraic equations
\bea
\label{eqpureFP1}
v_0&=&\frac{1}{128\pi^2(\gamma-f_0)^2}\left[ 
(N-10)\gamma^2-2(N-13)\gamma f_0+(N-4)f_0^2
-12\gamma^2\frac{\gamma-2f_0}{\gamma-f_0}\log\left(\frac{f_0}{\gamma}\right) \right]
\nonumber\\
0&=& -2f_0-\frac{4N-55}{384\pi^2}
-\frac{f_0(\gamma+9f_0)}{96\pi^2(\gamma-f_0)^2}
-\frac{\gamma\left(2\gamma^2-6\gamma f_0+9f_0^2\right)
\log\left(\frac{f_0}{\gamma}\right)}{48\pi^2(\gamma-f_0)^3}
\eea
These can be solved numerically.
The linearization of the flow equation around FP1 yields the following
equations:
\bea
0&=& 
-(\lambda+4)\delta v 
+\left(\varphi-\frac{N-1}{32\pi^2\varphi}\right)\delta v'
\nn\\
&&
-\left(\frac{3\gamma\left(3f_0^2+5\gamma f_0-2\gamma^2\right)}
{16\pi^2f_0(f_0-\gamma)^3}
+\frac{3 \gamma ^2 (\gamma -4f_0)}{8\pi^2(f_0-\gamma)^4}\log\left(\frac{f_0}{\gamma}\right)\right)\delta f
-\frac{1}{32\pi^2}\delta v''
\\
0&=& -(\lambda +2)\delta f
+\left(\frac{\gamma\left(37 f_0^2-11\gamma f_0+4\gamma^2\right)}
{96\pi^2 f_0(f_0-\gamma)^3}
-\frac{\gamma f_0(2\gamma+3f_0)}{16\pi^2(f_0-\gamma)^4}
\log\left(\frac{f_0}{\gamma }\right)\right)\delta f
+\left(\varphi-\frac{N-1}{32\pi^2\varphi}\right)\delta f'
\nn\\
&&
+\frac{N-1}{96\pi^2\varphi}\delta v'
+\frac{1}{96\pi^2}\delta v''
-\frac{1}{32\pi^2}\delta f''
\eea

For $N=1$ and $\gamma=1$ we have $v_{0*}=0.03314$, $f_{0*}=0.01552$.
The relevant eigenperturbations around this solution 
can be found numerically and the corresponding eigenvalues are
$-4$, $-2.272$, $-2$, $-0.272$, $0$.
There are therefore four relevant and one marginal couplings.
For $\gamma=0.006$ we have $v_{0*}=0.00353$, $f_{0*}=0.00670$,
which are very close to the values found with the other cutoff in \cite{pv1}.
The relevant eigenperturbations around this solution 
have eigenvalues $-4$, $-2.542$, $-2$, $-0.542$, $0$,
which are also closer the the other cutoff.

For $N=4$ and $\gamma=1$ the fixed point is at $v_{0*}=0.03635$, $f_{0*}=0.01413$.
The linear perturbation analysis around the fixed point gives the following eigenvalues: $-4$, $-2.299$, $-2$, $-0.299$, $0$.
If instead we choose $\gamma=0.006$, we have $v_{0*}=0.00669$, $f_{0*}=0.00549$,
and the eigenvalues are $-4$, $-2.682$, $-2$, $-0.682$, $0$.

In this scalar-free cutoff scheme it is interesting to investigate the solutions of Eqs. ~\eqref{eqpureFP1} for FP1 in the large $N$ limit.
It is easy to see that in this limit they become
\be
v_0\approx \frac{16N-205}{512\pi^2} \quad , \quad 
f_0\approx \gamma\exp{\frac{55-4N}{16}}\ .
\ee
We see that this behaviour is very different from the one obtained previously: in particular $f_0$ becomes exponentially small
but never changes sign.
Thus there is apparently no upper bound in $N$ from the requirement of having a positive $f_0$. This behavior is induced by the interplay between the $\log$ singularity in $f_0$ and the linear dependence in $N$ in the equations.
This result is probably not physically correct for the following 
reason: in the ERGE one should use the Hessian defined as the second derivative of the Effective Average Action (EAA) with respect to the quantum field.
Instead in order to close the flow equation, 
we are using the second derivative of the
EAA with respect to the background field.
Even though the function $F$ does not appear in the cutoff,
it does appear in the denominator, where the coefficient
of $-\nabla^2$ is $F-\gamma k^{d-2}$.
This term is absent with the type-I cutoff and
it is its presence with the scalar-free cutoff that gives rise
to the logarithmic terms in the flow equation.
In a proper bi-metric calculation the coefficient of 
$-\nabla^2$ in the denominator of the ERGE
would be $Z_h-\gamma k^{d-2}$, where $Z_h$ is the graviton
wave function renormalization constant.
The argument of the logarithmic term would then be $Z_h/\gamma$
and the beta function of $f$ would probably be regular for $f=0$.

\section{Discussion}

This paper extends earlier investigations of scalar-tensor
gravity, where we used the exponential parametrization of the metric.
The {\it a priori} motivation of this parametrization is that
it respects the nonlinear structure of the space of riemannian metrics.
It appears a posteriori that it leads to better behaved flows:
there are no infrared singularities \cite{pv1,falls}
and the equations can be made gauge-independent even off-shell
\cite{falls}.
Quite generally, the use of the exponential parametrization
in conjunction with the physical gauge $\xi_\mu'=\nabla_\mu h=0$
leads to simpler equations.
In the theory considered here, they are sufficiently simple
that analytic scaling solutions can be found.
We have given here the form of such solutions in any dimensions
and for any number of scalar fields.
Relative to our earlier work, the advantage of having more than
one scalar field is that there exists a global solution (FP4)
with $f\geq 0$ everywhere.

One interesting by-product of this analysis is the confirmation
of upper bounds on the number of scalars, for compatibility with
a fixed point with positive $f$.
This result is not confirmed when one uses a scalar-free cutoff
but we have argued that this is probably not physically correct.

As is often the case when taking the large $N$ limit we have been able to construct the exact analytical solutions
and show that they are unphysical in the type-I cutoff scheme.

We have also investigated in $d=3$ the existence of the gravitationally dressed Wilson-Fisher scaling solution for various finite N.
In particular we have shown that physically acceptable solution do exist for small $N$ and given details for the case $N=2$. 

We hope that these tools will prove useful also in the search
of global solutions for $f(R)$ truncations,
where several results have been obtained recently
\cite{various}, but a clear picture is still missing.
Another important direction of investigation is the inclusion of other matter fields, in particular fermions~\cite{Yuk,EG,VZ}.
If the picture we have found up to now will be confirmed also in more sophisticated truncations,
the possibility to define a consistent QFT of matter-gravity interactions will be a guidance in the understanding of several aspects of fundamental/effective physics.
We find already interesting to start to investigate possible implications at phenomenological level for example in cosmology. 

{\bf Acknowledgements.}
We would like to thank Astrid Eichhorn and Jan Pawlowski
for discussions.

\appendix

\section{Hessian}

The Hessian for gravity coupled to a single scalar has been
given in \cite{pv1}.
Here we discuss the changes introduced by having a scalar multiplet for the model of Eq.~(\ref{action}).
The scalar field is decomposed into a background and 
fluctuation part 
$\phi^a=\bar{\phi}^a+\delta\phi^a$.
The part of the Hessian which contains a dependence 
on the scalar fluctuation $\delta\phi^a$ is a sum of two pieces.
The first is quadratic in the scalar field fluctuations and reads:
\bea
\int \mathrm{d}^d x \sqrt{\bar{g}}\; \frac{1}{2} \delta\phi^a
\bigg[ 
\left(-\bar{\nabla}^2
+\bar V''-\bar F''\bar{R}\right) P^{R}_{ab} 
+\left(-\bar{\nabla}^2
+\frac{1}{\bar\rad}(\bar V'-\bar F'\bar{R})\right) P^{\perp}_{ab}
\bigg] \delta\phi^b
\eea
where we have introduced the
projectors $P_R^{ab}=\bphi^a\bphi^b/\bphi^2$ 
and $P_\perp^{ab} =\delta^{ab}-P_R^{ab}$.
The second mixes scalar field fluctuations with metric scalar fluctuations:
\be
\int \mathrm{d}^d x \sqrt{\bar{g}}\; 
\delta\phi^a P^{R}_{ab} \bar{\phi}^b
\frac{1}{\bar\rad}
\bigg[\frac{1}{2}h(\bar V'-\bar F'\bR)-\bar F'(\bar \nabla_\mu\bar \nabla_\mu h^{\mu\nu} - \bar{\nabla}^2 h - \bar{R}_{\mu\nu} h^{\mu\nu}) \bigg]
\ee
Without loss of generality we can separate the radial field component from the Goldstone bosons by fixing the background to be
$\bar{\phi}^a = 0,\;\;a=1,\dots,N-1$, $\bar{\phi}^N = \bar{\varphi}$
and define $\delta\phi^N = \delta \varphi$. 
After the York decomposition
\be
\label{york}
h_{\mu\nu}=\tth_{\!\!\mu\nu}
+\bnabla_\mu\xi_\nu
+\bnabla_\nu\xi_\mu
+\bnabla_\mu\bnabla_\nu\sigma
-\frac{1}{d}\bg_{\mu\nu}\bnabla^2\sigma +\frac{1}{d}\bg_{\mu\nu} h
\ee
and the redefinition
\be
\label{redef}
\xi'_\mu=\sqrt{-\bar\nabla^2-\frac{\bar R}{d}}\xi_\mu\ ;\qquad
\sigma'=\sqrt{-\bar\nabla^2}\sqrt{-\bar\nabla^2-\frac{\bar R}{d-1}}\sigma\ .
\ee
 the second piece can be rewritten
\be
\int \mathrm{d}^d x \sqrt{\bar{g}}\; 
\frac{(d-1)}{d}\bar F' \delta\varphi
\bigg[ -\sqrt{-\bar{\nabla}^2\left(-\bar{\nabla}^2-\frac{\bar{R}}{d-1}\right)} \, \sigma' 
+\left(-\bar{\nabla}^2
-\frac{(d-2) \bar{R}}{2(d-1)}
+\frac{d\,\bar{V}^{\prime}}{2(d-1)\bar{F}^{\prime}}\right)h\bigg]
\ee
Fixing the gauge $h=const$, $\xi=0$ and also neglecting the residual constant mode of $h$, the full Hessian becomes
\bea
\Gamma_k^{(2)} = \int \mathrm{d}^d x \sqrt{\bar{g}} \; \bigg\lbrace
\frac{1}{4}\,\bar{F}\, h^{TT}_{\mu\nu} \left( -\bar{\nabla}^2 + \frac{2 \bar{R}}{d(d-1)} \right) h^{TT\,\mu\nu}
- \frac{(d-2)(d-1)}{4\,d^2}\,\bar{F} \sigma^{\prime} (-\nabla^2) \sigma^{\prime} 
\nonumber\\
+\frac{1}{2}\delta\varphi(-\bar{\nabla}^2+\bar V''-\bar F''\bar{R})\delta\varphi
+\frac{1}{2}\sum_{a=1}^{N-1}\delta\phi^a 
\left(-\bar\nabla^2
+\frac{1}{\bar\rad}(\bar V'-\bar F'\bar{R})\right)\delta\phi^a \label{oldHessian} 
\\ \nonumber
- \,   \frac{(d-1)}{d}\bar{F}^{\prime}\delta\varphi
 \sqrt{-\bar{\nabla}^2\left(-\bar{\nabla}^2-\frac{\bar{R}}{d-1}\right)} \, \sigma^{\prime} \bigg\rbrace\,.
\eea
The Hessian can then be diagonalised by the transformation
\be
\sigma^{\prime\prime} = \sigma^{\prime} 
-\frac{2d}{d-2}\frac{\bar{F}^{\prime}}{\bar{F}} \sqrt{\frac{-\bar{\nabla}^2-\frac{\bar{R}}{d-1}}{-\bar{\nabla}^2}} \, \delta\varphi\,,
\ee
and the diagonal Hessian reads
\bea
\label{hessfinal}
\Gamma_k^{(2)} = \int \mathrm{d}^d x \sqrt{\bar{g}} \bigg\lbrace
\frac{1}{4}\,\bar{F}\, h^{TT}_{\mu\nu} \left( -\bar{\nabla}^2 
+ \frac{2 \bar{R}}{d(d-1)} \right) h^{TT\,\mu\nu}
- \frac{(d-2)(d-1)}{4\,d^2}\,\bar{F} \sigma^{\prime\prime} (-\nabla^2) \sigma^{\prime\prime} 
\nonumber\\
+\, \frac{1}{2} \delta\varphi \bigg[ 
-\bar{\nabla}^2+\bar V''-\bar F''\bar{R}
+ 2\frac{d-1}{d-2}\,\frac{\bar{F}^{\prime 2}}{\bar{F}} \left( -\bar{\nabla}^2 
- \frac{\bar{R}}{d-1} \right)
\bigg] \delta\varphi\\ \nonumber
+\,\frac{1}{2}\sum_{a=1}^{N-1}\delta\phi^a 
\left(-\bar\nabla^2
+\frac{1}{\bar\rad}(\bar V'-\bar F'\bar{R})\right)\delta\phi^a  \bigg\rbrace\,.
\eea

The ghost terms induced by the gauge fixing are given by:
\be
\label{ghosts}
S_{gh:h}=\int d^dx\sqrt{\bar g}\, c(-\bar\nabla^2)c 
\quad , \quad
S_{gh:\xi}=\int d^dx\sqrt{\bar g}\,  c_\mu\bar g^{\mu\nu}\left(-\bar\nabla^2-\frac{\bar R}{d}\right)c_\nu\,.
\ee



\begin{thebibliography}{99}

\bibitem{W1}
S.~Weinberg,
in Hawking, S.W., Israel, W.: General Relativity (Cambridge University Press), 790-831.

\bibitem{pv1}
  R.~Percacci and G.~P.~Vacca,
  Eur.\ Phys.\ J.\ C {\bf 75} (2015)  188
  [arXiv:1501.00888 [hep-th]].

\bibitem{hprw}
T. Henz, J. Pawlowski, A. Rodigast and C. Wetterich,
Phys. Lett. B727 (2013) 298-302
arXiv:1304.7743 [hep-th]

\bibitem{cpr2}
A.~Codello, R.~Percacci and C.~Rahmede,
Annals Phys.\  {\bf 324} (2009) 414
[arXiv:0805.2909 [hep-th]].

\bibitem{Litim}
D.~F.~Litim,
Phys.\ Rev.\ D {\bf 64} (2001) 105007
[hep-th/0103195].

\bibitem{dep}
P. Dona', A. Eichhorn and R. Percacci,
Phys.Rev. D89 (2014) 084035
arXiv:1311.2898 [hep-th]

\bibitem{marchais}
E. Marchais, Ph.D. thesis, University of Sussex.

\bibitem{narain1}
G. Narain, R. Percacci, Class. Quant. Grav. 27 (2010) 075001,
arXiv:0911.0386[hep-th]

\bibitem{bk}
J. Borchardt and B. Knorr,
arXiv:1502.07511 [hep-th]

\bibitem{dietz3}
I.Hamzaan Bridle, J. A. Dietz and Tim R. Morris
JHEP 1403 (2014) 093 
arXiv:1312.2846 [hep-th]

\bibitem{narain3}
G. Narain and R. Percacci,
Acta phys, Polon. B40 3439-3457 (2009),
arXiv:0910.5390 [hep-th]

\bibitem{dietz4}
J.A. Dietz, T.R. Morris, 
arXiv:1502.07396 [hep-th]

\bibitem{becker}
D. Becker and M. Reuter,
Annals of Phys. 350 (2014) 225-301,
arXiv:1404.4537 [hep-th]

\bibitem{falls}
K. Falls,
arXiv:1501.05331 [hep-th]

\bibitem{various}
D. Benedetti and F. Caravelli,
JHEP 1206 (2012) 017, Erratum-ibid. 1210 (2012) 157,
arXiv:1204.3541 [hep-th]

D.~Benedetti,
Europhys.\ Lett.\  {\bf 102} (2013) 20007,
[arXiv:1301.4422 [hep-th]].

M. Demmel, F. Saueressig and O. Zanusso,
JHEP 1211 (2012) 131,
arXiv:1208.2038 [hep-th]

J.A. Dietz and T.R. Morris,
JHEP 1301 (2013) 108,
arXiv:1211.0955 [hep-th]

J.A. Dietz and T.R. Morris,
JHEP 1307 (2013) 064,
arXiv:1306.1223 [hep-th]

M. Demmel, F. Saueressig and O. Zanusso,
JHEP 1211 (2012) 131,
arXiv:1504.07656 [hep-th]

\bibitem{Yuk}
  O.~Zanusso, L.~Zambelli, G.~P.~Vacca and R.~Percacci,
  Phys.\ Lett.\ B {\bf 689} (2010) 90
  [arXiv:0904.0938 [hep-th]];
  G.~P.~Vacca and O.~Zanusso,
  Phys.\ Rev.\ Lett.\  {\bf 105} (2010) 231601
  [arXiv:1009.1735 [hep-th]].
 
 \bibitem{EG}
  A.~Eichhorn and H.~Gies,
  New J.\ Phys.\  {\bf 13} (2011) 125012
  [arXiv:1104.5366 [hep-th]].

 \bibitem{VZ}
  G.~P.~Vacca and L.~Zambelli,
  arXiv:1503.09136 [hep-th], to appear on PRD.
  
\end{thebibliography}
\end{document}